\documentclass[
    sort&compress
&&  ,final            
    ,draft            
    ,numberedheadings 
  ]
  {aipproc}

\layoutstyle{6x9}

\usepackage{aipxfm}
\usepackage{pubs}
\usepackage{amsmath}
\usepackage{amssymb}
\newcommand{\ud}{\mathrm{d}}
\begin{document}

\title[ ]{Simulation of prompt emission from GRBs with a
photospheric component and its detectability by GLAST}


\author{Milan Battelino}{
  address={Stockholm Observatory, AlbaNova University Center, 106 91 Stockholm, Sweden}
}

\author{Felix Ryde}{
  address={Stockholm Observatory, AlbaNova University Center, 106 91 Stockholm, Sweden}
}

\author{Nicola Omodei}{
  address={INFN Pisa, Largo B.Pontecorvo 3, 56100 Pisa, Italy}
}

\author{Francesco Longo}{
  address={University and INFN of Trieste, via Valerio 2, 34100 Trieste, Italy}
}

\begin{abstract}
The prompt emission from gamma-ray bursts (GRBs) still requires a
physical explanation.  Studies of time-resolved GRB spectra,
observed in the keV-MeV range, show that a hybrid model consisting
of two components, a photospheric and a non-thermal component, in
many cases fits bright, single-pulsed bursts as well as, and in
some instances even better than, the Band function. With an energy 
coverage from 8 keV up to 300 GeV, GLAST will give
us an unprecedented opportunity to further investigate the nature
of the prompt emission. In particular, it will give us the
possibility to determine whether a photospheric component is the
determining feature of the spectrum or not. Here we present a short
study of the ability of GLAST to detect such a photospheric
component in the sub-MeV range for typical bursts, using
simulation tools developed within the GLAST science collaboration.
\end{abstract}

\maketitle


\section{Introduction}
The Band function \citep[see][]{Bandetal1993}, a softly broken power-law, 
describes the keV-MeV spectra for a broad range of gamma-ray bursts (GRB) 
exceptionally well. It is an
empirical function without any claim to describe the physical
processes behind the continuum spectra. However, it has properties
that in many bursts could be described as the result of
non-thermal radiation processes, such as synchrotron or inverse
Compton scattering. Some spectra, even though well-described by
the Band function, can not be the result of plain
\emph{optically-thin synchrotron} (OTS) processes since the photon
index of the sub-peak spectral slope have a value $\alpha$ larger
than $-2/3$, and hence lies beyond the synchrotron "line of
death", LOD \citep[see][]{Crideretal1997, Preeceetal1998}.
Synchrotron processes may however still be a viable option through
realization of {\it synchrotron self-absorption} (SSA) in a higher
energy domain, see e.g. \citet{LloydPetrosian2000}. The physical
conditions required to produce such hard spectral slopes in the
100 keV domain are generally considered to be unreasonable but
too little is known regarding the generation of magnetic
fields and the relevant conditions in order to completely rule out this
possibility. Inverse Compton scattering with a self-absorbed
synchrotron seed spectrum residing in the optical domain  ({\it
synchrotron self-Compton}, SSC) is a possible mechanism for
spectra with $\alpha$ larger or equal to $0$, see
\citet{PanaitescuMeszaros2000}. Still another explanation for the
hard spectral slopes, suggested by \citet{Medvedev2000}, is a
combination of jitter radiation and OTS radiation. \citet{BaringBraby2004}
argued, however, that since both the latter emission processes
need an almost purely non-thermal electron distribution to be able
to fit the observed spectra, these emission models are difficult
to reconcile with the assumed diffusive shock acceleration
models, which often give rise to a strong contribution of a
thermal population.

Models predicting a photospheric emission component in the prompt
spectra of GRBs has been suggested by several authors
\citep[see][]{MeszarosRees2000, Meszarosetal2002,
DaigneMochkovitch2002, DrenkhahnSpruit2002, Ryde2004a, Ryde2005,
ReesMeszaros2005, Rydeetal2006}. \citet{MeszarosRees2000} argue
that the contribution of thermal radiation -- originating from the
expanding fireball when it becomes optically thin to Thomson
scattering -- could explain the hard spectral slopes in the prompt
GRB spectra. Ryde shows in \cite{Ryde2004a} that some bright {\it
CGRO} BATSE bursts have sub-peak spectral slopes that fit
blackbody radiation very well, in agreement with results by 
e.g. \citet{Ghirlandaetal2003}, and also demonstrates in \cite{Ryde2005}
that time-resolved BATSE spectra can, as an alternative to the
phenomenological Band function, be fitted with a hybrid model
consisting of a single power-law and a blackbody function. The
hybrid model gives, for bright single pulsed bursts, just as good
fit and in many cases even a better fit than the Band function.
Hence, by combining a photospheric component with a power-law
function, that can be interpreted as being a result of further
energy dissipation in the optically thin part of the outflow, a
wide range of GRBs observed by BATSE can be explained.

In this paper, we investigate the GLAST detectability of bursts
containing a strong photospheric component in the sub-MeV range. A
sample of 847 time-resolved spectra from 57 bright BATSE bursts
were analyzed where both the Band function and the hybrid model was
imposed on the data. The fits and analysis were performed with
XSPEC \citep{ArnaudDorman2003}. Three representative bursts were
selected for further detailed analysis and used as basis for
simulations in the expanded energy range of the two instruments 
onboard GLAST: the GLAST Burst Monitor (GBM), covering an energy 
range from 8 keV to 30 MeV (hence encompassing energies both below 
and above the BATSE window), and the Large Area Telescope (LAT), 
including energies from 20 MeV up to 300 GeV. The three
selected bursts, GRB911016, GRB941026 and GRB960530, were all
single-pulsed, had comparable $\chi^2$-values and "goodness-of-fit" 
($P$-values) for both models, and showed a low-energy power-law index (Band function) that
lied beyond LOD, i.e. with $\alpha$ larger than $-2/3$. The
temporal features associated with each of the three bursts were
extracted and the resulting data extended from the BATSE energy
range (20 keV - 2 MeV) into the GLAST domain (8 keV - 300 GeV), hence assuming a strong high-energy 
emission. The high-energy part is generally considered to be characterized 
by the Band function spectral index $\beta$ that for the majority of BATSE bursts has a value 
less than $-2$. Of the few observed bursts that has high-energy spectra 
available, some do however show a strong high-energy emission \cite{Gonzalezetal2003}. 
In section \ref{sec:modeling} we first describe how the BATSE 
data was parameterized and the model that was used for the GLAST 
simulations. Section \ref{sec:simulating} explains how the simulated 
GLAST data was produced and section \ref{sec:results} presents 
the results. We conclude this paper with a discussion in section 
\ref{sec:discussion}.

\section{Modeling the thermal GRBs}
\label{sec:modeling}
\subsection{The hybrid model}
\begin{figure}[htb]
  \includegraphics[width=1.0\textwidth]{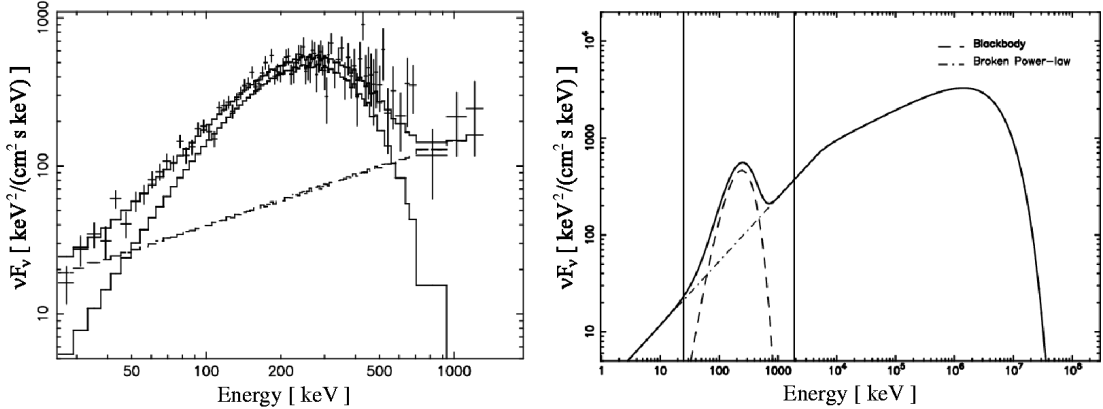}
  \caption{Time-resolved $\nu F_{\nu}$ spectra of GRB911016 assumed to 
     have a strong high-energy component in the GeV-domain. See also
     figure \ref{fig:GRB911016_GLAST_ufu_chi2}.
     \emph{Left panel:} XSPEC Hybrid model fit in BATSE energy band: 
     0.704 - 1.280 s. 
     \emph{Right panel:} SBM hybrid model simulation of the same burst 
     extrapolated in the GLAST energy band: 1.0 - 1.1 s. The solid line 
     describes the hybrid model, the dashed line the blackbody component 
     and the dot-dashed line the broken power-law. The two vertical lines 
     in the plot describe the energy range covered by BATSE.}
   \label{fig:grb911016_batse_sbm}
\end{figure}

The hybrid model consists of two components, a photospheric and a
non-thermal radiation spectrum. These were in our analysis and in 
the subsequent simulations represented by a blackbody spectrum 
(Planck function) and a single power-law, respectively. This model 
is in XSPEC described by $N_{hybrid}(E)=N_{bb}(E)+N_{po}(E)$ where:
\begin{eqnarray}
\label{eqn:Nbb}
N_{bb}(E) &=& A_{bb}\,\frac{E^2\ud E}{e^{E/kT}-1} \\
\label{eqn:Npo} N_{po}(E) &=& A_{po}\,\left( \frac{E}{1\,keV}
\right)^{-p}
\end{eqnarray}
$A_{bb}$ and $A_{po}$ are normalization constants for respective
 component, $kT$ the temperature in keV for the blackbody component
and $p$ the photon index of the power-law component. This simple
two-component model was used when analyzing the BATSE data in the
20 keV-1 MeV energy range. The left panel in figure
\ref{fig:grb911016_batse_sbm} shows a hybrid model fit to a
time-resolved spectrum of GRB911016 covering the time-interval
0.704-1.280 s after the trigger.
In our simulations for GLAST we extended the energy range into the
GLAST domain, and therefore used an extended time-dependent hybrid
model that included a blackbody function that evolves over time
and a broken power-law -- instead of a single power-law component
-- with a high-energy cut-off. The blackbody component that
evolves over time is described by:
\begin{equation}
\label{eqn:Nbb2}
N_{bb}(E,t) \,=\, A(t)_{bb}\,\frac{E^2\ud E}{e^{E/kT(t)}-1}
\end{equation}
We motivated the broken power-law based on the following two
facts: {\it a)} OTS radiation may according to theory show a break
in the energy domain covered by GLAST and {\it b)} power-law index
values in the BATSE data from the hybrid model fits can be
interpreted as being distributed around two values as shown in
\cite{Battelino2006}.
\begin{table}
\caption{Power-law photon indices extracted with BATSE and used as
input to SBM}
\begin{tabular}{cccccc}
\hline
    \tablehead{2}{c}{b}{GRB911016}
  & \tablehead{2}{c}{b}{GRB941026}
  & \tablehead{2}{c}{b}{GRB960530} \\
    \tablehead{1}{c}{b}{Time range [s]}
  & \tablehead{1}{c}{b}{Index}
  & \tablehead{1}{c}{b}{Time range [s]}
  & \tablehead{1}{c}{b}{Index}
  & \tablehead{1}{c}{b}{Time range [s]}
  & \tablehead{1}{c}{b}{Index}        \\
\hline
0.000 - 3.520 & $1.30\pm0.05$  & 0.030 - 6.272  & $1.57\pm0.09$ &  0.029 - 18.624 & $1.42\pm 0.07$ \\
3.520 - 7.105 & $1.69\pm0.08$  & 6.272 - 14.784 & $2.12\pm0.04$ &        -        &       -   \\
\hline
\end{tabular}
\end{table}
These two power law index values may be given a physical
interpretation: the higher value ($\sim 2.1$) stems from a
shock-accelerated distribution of electrons and the lower value
($\sim 1.5$) from the cooling electrons originating from the same
distribution. The broken power-law component is described by:
\begin{eqnarray}
N(E,t)_{bpo}\,=\,
\begin{cases}
A(t)_{bpo}\,(\frac{E}{1\,keV})^{p_1}, \qquad \qquad \qquad \qquad \, \, \, \, E\,<\,E_b(t) \\
A(t)_{bpo}\,(\frac{E_b(t)}{1\,keV})^{p_2-p_1}\,(\frac{E}{1\,keV})^{-p_2}, \qquad \quad E\,\geq\,E_b(t) \\
\end{cases}
\end{eqnarray}
A high-energy cut-off was assumed due to the competition between
the acceleration of the electrons and the radiative cooling that leads
to a maximal energy that the electrons can be accelerated to. The
values chosen for the simulations were based on results by
\citet{deJageretal1996}. Their studies show that the spectrum of the
Crab nebula has a high-energy cut-off $e$-folding energy of
$\sim$30 MeV, as predicted by their model. This is a robust value
given by the model and independent of the magnetic field strength,
$B$. We assumed, in our hybrid model implementation, that similar
physical conditions exist at the GRB radiation site.  Since the
spectra from the prompt emission is boosted with a Lorentz-factor
$\Gamma \sim 100$, we selected $e$-folding values of 5.0 GeV for
GRB911016 and 3.0 GeV for GRB941026 and GRB960530. These values
fall within the GLAST energy range and the extended hybrid model
is therefore in the simulations described by:
\begin{eqnarray}
\label{eqn:exthybrid}
N(E,t)\,=\,
\begin{cases}
N_{bb}(E,t)\,+\,N_{bpo}(E,t), \qquad \qquad \qquad \,\,\,\,E\,<\,E_{c} \\
N_{bb}(E,t)\,+\,N_{bpo}(E,t)\,e^{(E_{c}-E)/E_{e}}, \qquad \,\,\,E\,\geq\,E_{c}\\
\end{cases}
\end{eqnarray}
where $E_{c}$ is the energy from where the high-energy cutoff
starts and $E_{e}$ the $e$-folding energy. {\it Extra-galactic
background light} (EBL) is expected to contribute to a high-energy
cut-off for GRBs, but we did not consider this $\gamma \gamma$
absorption process in our simulations. This absorption process may
however be important in spectra from GRBs at large cosmical
distances.

\subsection{Modeling the temporal evolution}
\citet{Kocevskietal2003a} show that the FRED light-curves
associated with the temporal evolution of the flux of single pulse
bursts may be described by:
\begin{equation}
\label{eqn:lcurve}
F(t)\,=\,F_m\bigg(\,\frac{t}{t_m}\,\bigg)^r\,\bigg[\frac{d}{d+r}\,+\,\frac{r}{d+r}\,\bigg( \frac{t}{t_m} \bigg)^{r+1} \bigg]^{-(r+d)/(r+1)}
\end{equation}
where $F_{m}$ is maximum flux, $t_m$ the time at maximum flux, $r$
the dimensionless power-law index describing the rise phase and
$d$ the corresponding index for the decay phase. The power-law
describing the rise phase is in equation (\ref{eqn:lcurve})
proportional to $t^r$ while the decay phase is described by:
\begin{equation}
F(t)\,=\,F_0\,\big[ 1+\frac{t}{T(d-1)} \big]^{-d}
\end{equation}
as suggested by \citet{RydeSvensson2000,RydeSvensson2002} for
single pulse bursts. $F_0$ is here the flux at the start of the
decay phase and $T$ a normalization parameter. By utilizing the
Levenberg-Marquardt algorithm we extracted the $r$ and $d$
parameters using equation (\ref{eqn:lcurve}) for the BATSE flux of
each model component (blackbody and single power-law) over the
energy range 20 keV-1 MeV for the three bursts.

\begin{table}
\caption{Temperature parameters used as input to SBM}
\label{tab:temperature}
\begin{tabular}{l c | c  c  c}
\hline
\tablehead{2}{c}{b}{}
  & \tablehead{1}{c}{b}{GRB911016}
  & \tablehead{1}{c}{b}{GRB941026}
  & \tablehead{1}{c}{b}{GRB960530}   \\
\hline
 $kT_n$    & [keV] & 67.23          & 67.6            & 37.38           \\
 $t_n$    & [s]   & 0.992          & 0.960           & 2.392           \\
 $t_0$    & [s]   & $1.24\pm0.06$  & $2.1\pm 0.2$    & $3.0\pm 0.3$    \\
 a        &       & $-0.04\pm0.03$ & $0.0$           & $-0.22\pm 0.05$ \\
 b        &       & $-0.48\pm0.02$ & $-0.35\pm 0.02$ & $-0.65\pm 0.04$ \\
 $\delta$ &       & $0.08$         & $0.2\pm0.1$     & 0.02            \\
\hline
\end{tabular}
\end{table}

Ryde shows in \cite{Ryde2004a} that $kT$ in equation
(\ref{eqn:Nbb2}) is a function of time that often follows a smoothly broken power-law with an
initial power-law index $a$ and a later index $b$:
\begin{equation}
\label{eqn:bbtemp}
kT(t)\,=\,kT_n\,\bigg( \frac{t}{t_n} \bigg)^\phi \, \bigg( \frac{ \cosh[ \log(t/t_0) / \delta ] } { \cosh [ \log (t_n/t_0) / \delta ] } \bigg)^{\xi \delta \ln 10}
\end{equation}
where $t_0$ is the time of the break, $\delta$ is the width of the
transition, $\xi\,=\,(b-a)/2$, and $\phi\,=\,(b+a)/2$. Equation
(\ref{eqn:bbtemp}) was used to extract the temporal parameters
associated with the blackbody component in the hybrid model. The
results from the fitting procedure, also here using a
Levenberg-Marquardt method, are presented in table
\ref{tab:temperature}.

The power-law indices, for all three bursts, exhibited some form
of temporal evolution, but it was -- due to the large variance for
each individual data point -- hard to determine whether the index
was smoothly changing over time or in discrete steps. One
interpretation of the varying power-law index is that the break in
synchrotron radiation spectrum -- that stem from a
shock-accelerated electron distribution (higher energy power-law
index) and from cooling electrons (lower energy power-law index)
-- moves or jumps across the BATSE energy window as the burst
progresses. For GRB911016 and GRB941026 we simulated the temporal
evolution of the power-law break with the following equation:
\begin{equation}
\label{eqn:break}
E_n(t)\,=\,(\,E_{n,0}\,-\,E_{n,\infty}\,)\,e^{-t/t_n}\,+\,E_{n,\infty}
\end{equation}
where $E_{n,0}$ is the initial energy at time $t=0$,
$E_{n,\infty}$ the asymptotic final energy when $t$ tends to
infinity and $t_n$ the rate scaling factor. This equation is
neither empirically nor analytically deduced, and is only
motivated by it being a candidate for describing the observed
evolution. 
The power-law component in GRB960530 could be well fitted with a
constant value and we did therefore not implement any temporal
evolution of its power-law break. We implemented, however, for
this simulation of GRB960530 a fixed break at 6.0 MeV, with a
high-energy power-law index of 2.1, as shown in figure
\ref{fig:GRB960530_hist}. 
\section{Simulating the thermal GRBs}
\label{sec:simulating} 
\begin{figure}[htb]
  \includegraphics[width=0.8\textwidth]{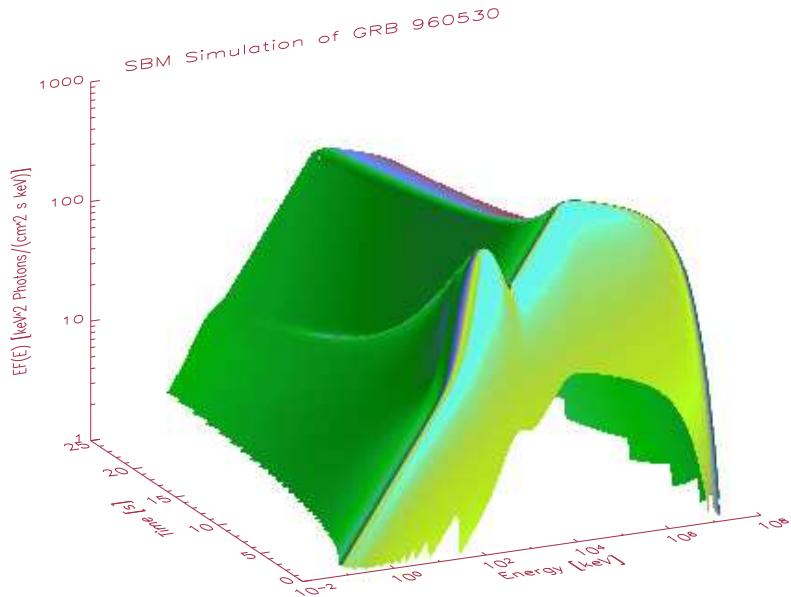}
  \caption{Evolution of the $\nu F_{\nu}$ spectrum
  over time in the keV-GeV range for a simulated GRB. The simulation is based
  on sub-MeV data collected by BATSE onboard CGRO for GRB960530. The blackbody
  peak can be seen around 100 keV, the power-law break around 10 MeV and the
  high-energy cutoff in the GeV domain.}
  \label{fig:GRB960530_hist}
\end{figure}

A generic C++ framework, Simple Burst
Modeler (SBM), was developed in order to simulate the spectral
evolution of the bursts covering the GLAST energy range. SBM
produces histogram files, one for each detector-type on-board
GLAST, that describe the amount of photons produced by the
simulated burst binned in energy and time. The right panel in
figure \ref{fig:grb911016_batse_sbm} show the time-resolved
spectrum of the simulated GRB911016 while figure
\ref{fig:GRB960530_hist} shows burst data simulated for GRB960530.
The histogram file produced for the LAT instrument is then fed
into {\it gtobssim}, the LAT fast observation simulator, and the
NaI and BGO histogram files into {\it GBM Tools}, the GBM
simulator. A modified version of {\it gtobssim} was used so that
generic histogram files of the type produced by SBM could be read
in and used in the simulations. When running {\it gtobssim} we
used the LAT instrument response file produced for {\it Data
Challenge 2} (DC2) in our simulations. {\it GBM Tools} was also
modified with the added capabilities to output the detector energy
grids into files required by SBM and to read in the histogram
files produced by SBM.
\begin{figure}[htb]
  \centering
  \includegraphics[width=0.8\textwidth]{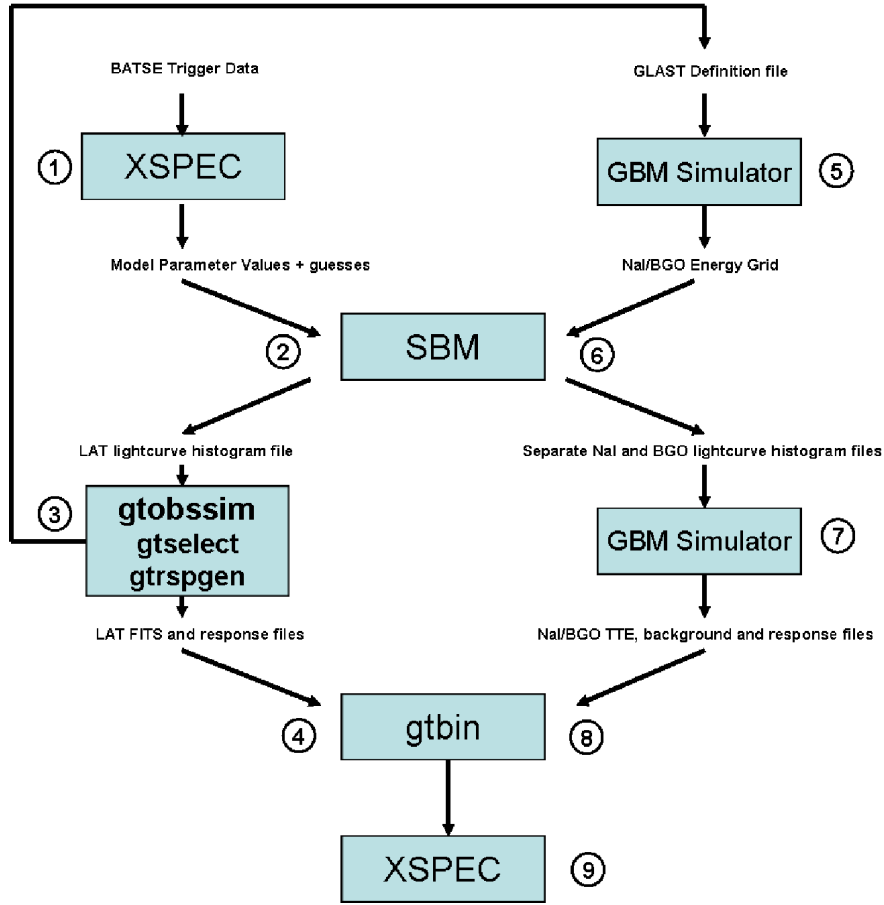}
  \caption{Combined SBM and GLAST simulation procedure. Note that
           \emph{gtobssim} has to be executed before the GBM Simulator since the definition file
           required by the GBM simulation package is created by the GLAST ScienceTools in
           connection to the LAT simulation run.}
\label{fig:anaproc}
\end{figure}
Figure \ref{fig:anaproc} summarizes the steps that were performed for each simulated burst. These seemingly awkward steps were needed since {\it gtobssim} produces a definition file used as input to GBM tools and GBM tools create energy grid files required by SBM.
\begin{enumerate}
  \item XSPEC was first used to analyze the BATSE data in order to extract the necessary parameters used as input to SBM.
  \item SBM was executed to produce a LAT histogram file.
  \item \emph{gtobssim} used the LAT histogram as input to produce the LAT FITS and response files, as well as the burst definition file. \emph{gtobssim} also needs a response function and a template that describes the position of the burst relative to the GLAST spacecraft. During our simulations the response function for DC2 was used. We also utilized the same template in all simulations hence positioning all three bursts at $\theta=62.9^\circ$ and $\phi=95.3^\circ$, where $\theta$ and $\phi$ represents the inclination angle to the normal of the LAT detector and azimuthal angle around the LAT normal respectively. \emph{gtselect} was used to select the events of interest and \emph{gtrspgen} to produce the response files readable by XSPEC.
  \item \emph{gtbin} was executed in order to create the FITS files in the PHA format readable by XSPEC. At this point all LAT data was produced.
  \item The GBM Simulator was now executed with the GLAST definition file as input, at this time only producing NaI and BGO energy grid files.
  \item SBM was executed again to produce histogram files that was used as input to the GBM simulator. The same parameter data was used as previously for the LAT run, but this time with the additional BGO and NaI energy grid files as input.
  \item The GBM Simulator was fed with the SBM histogram files producing {\it Time-Triggered Event} (TTE) files as well as background and response files.
  \item \emph{gtbin} was now utilized again but this time on the files produced by the GBM simulator converting them to the PHA format readable by XSPEC.
  \item The produced LAT data and GBM data were now jointly analyzed with XSPEC.
\end{enumerate}
\section{Results}
\begin{figure}[htb]
  \includegraphics[width=0.70\textwidth, angle=270]{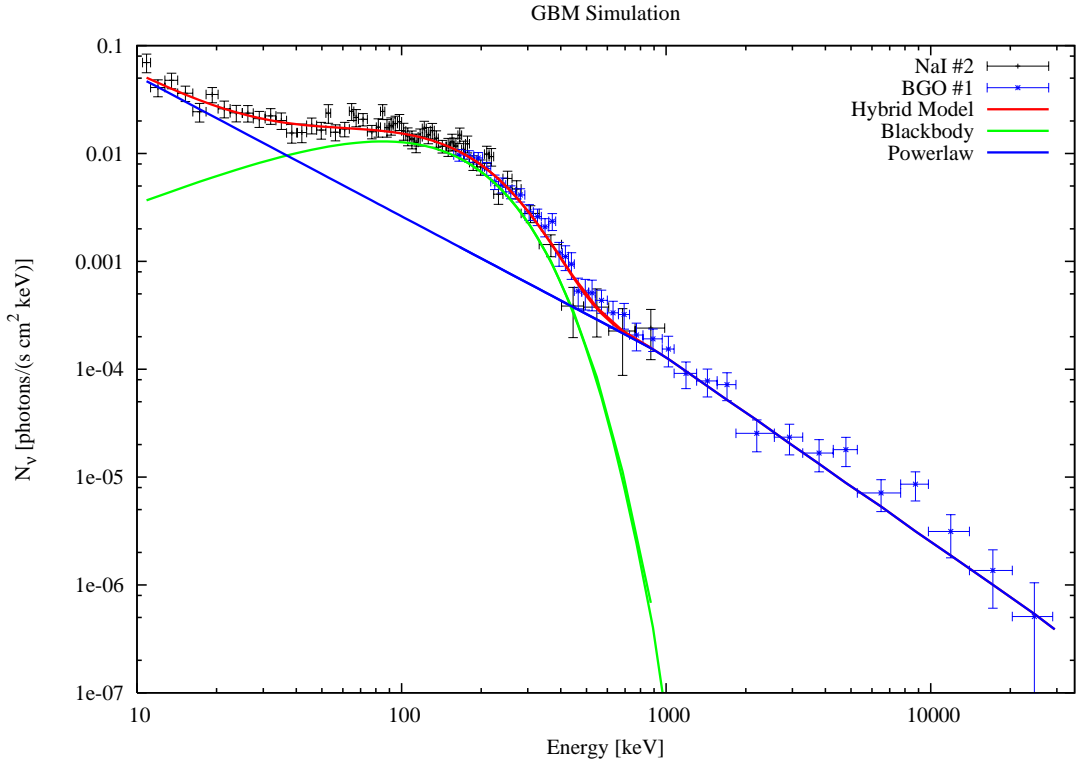}
  \caption{Unfolded photon-spectrum from one NaI and one BGO detector of the GBM instrument
  covering the first five seconds of the simulated burst. The fitted two components of
  the hybrid model are also shown in this figure. The fit was performed on ungrouped data
  using $\chi^2$ statistic. The simulation was based on BATSE data for GRB911016. Note that
  the data points have been grouped in the plot.}
  \label{fig:GRB911016_GLAST_efspec_s}
\end{figure}

\begin{figure}[htb]
  \includegraphics[width=0.70\textwidth, angle=270]{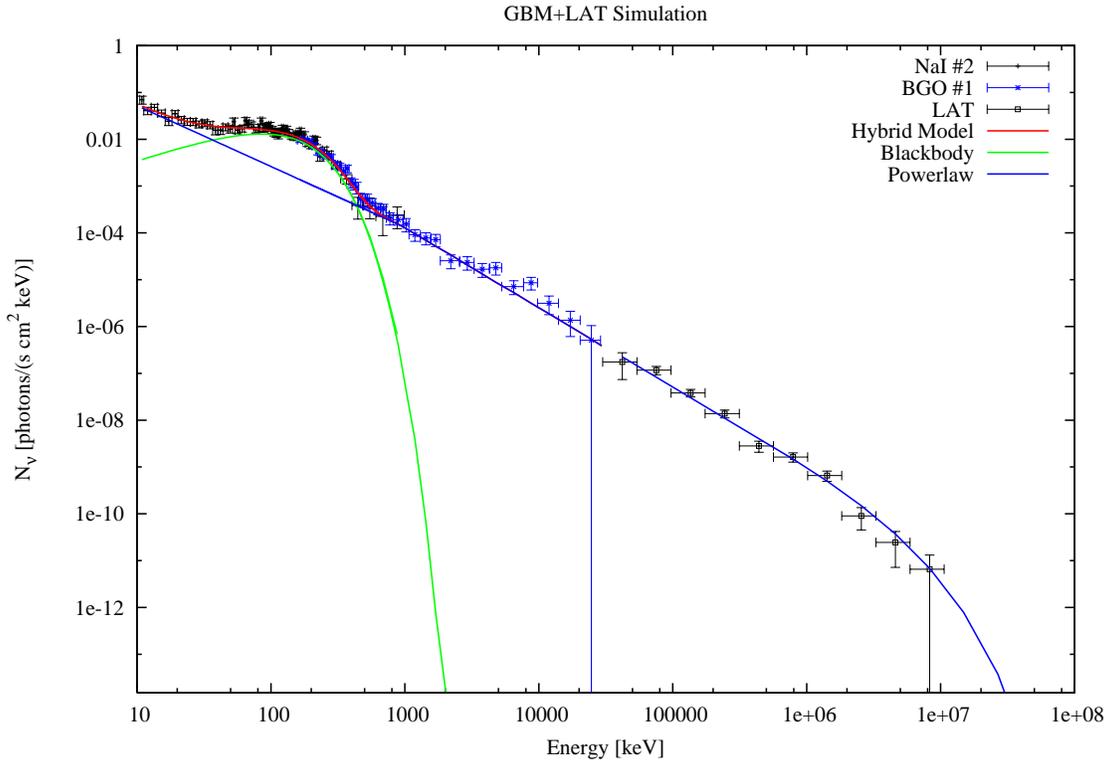}
  \caption{The same simulated burst as in figure \ref{fig:GRB911016_GLAST_efspec_s}, including 
           LAT data. The fit was performed on ungrouped data using $\chi^2$ statistic. Note 
	   that the data points has been grouped in the plot.}
  \label{fig:GRB911016_GLAST_efspec}
\end{figure}

\begin{figure}[htb]
  \includegraphics[width=1.0\textwidth]{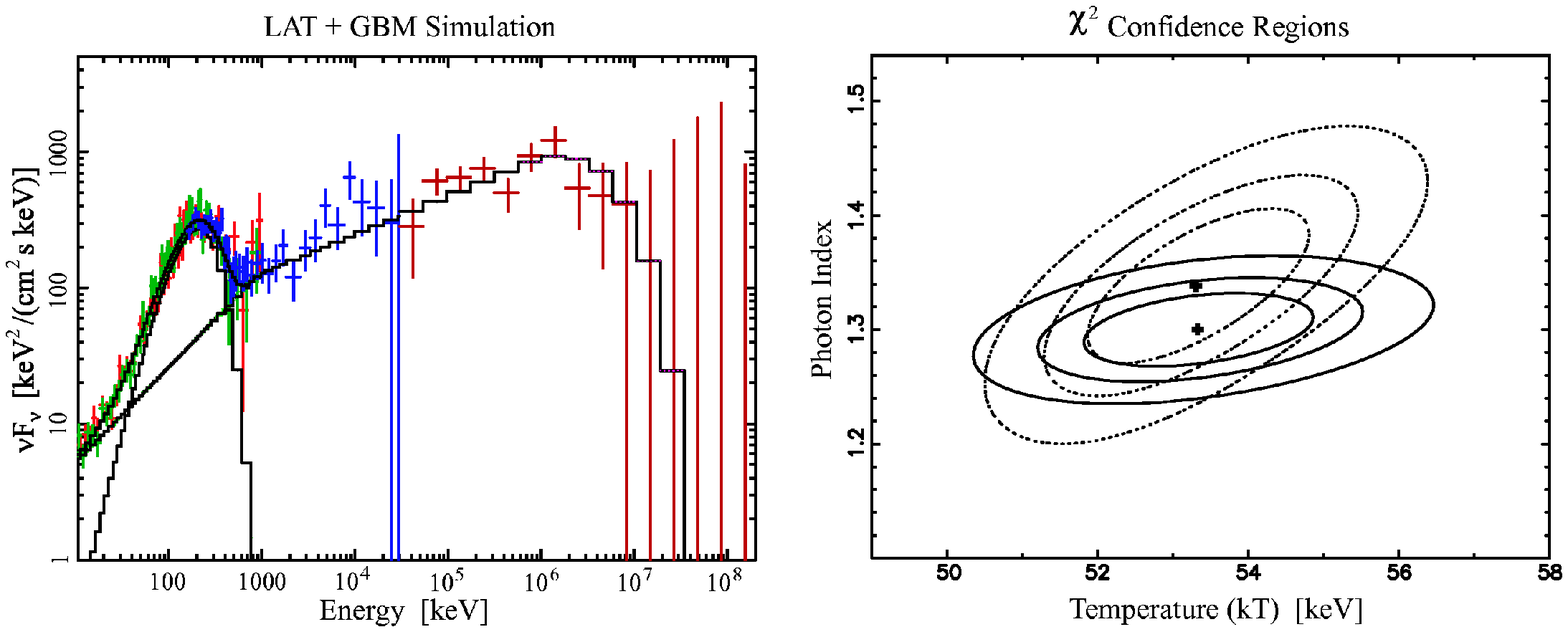}
  \caption{Burst data was simulated with SBM using parameters extracted
         from BATSE data for GRB911016. The SBM generated data was then used in the GBM and
         LAT observation simulations. \emph{Left panel:} $\nu F_\nu$ plot resulting from a fit
         of hybrid model from two NaI, one BGO detector and the LAT detector.
         This time-bin includes data from the first five seconds from the trigger.
         The fit was performed on ungrouped data using $\chi^2$ statistic.  Note
         that the data points has been grouped in the plot after the fit. See also figure 
         \ref{fig:grb911016_batse_sbm}.
         \emph{Right panel:} $\chi^2$ confidence regions for hybrid model fits with
         $1\sigma$, $2\sigma$ and $3\sigma$ levels. Photon index versus blackbody temperature
         for GRB911016 observed by BATSE (covering 0.029 - 4.864 s) have dotted contours and
     upper cross covering. The corresponding confidence regions
     for the simulated data for GLAST (covering 0.0 - 5.0 s) has solid contours and the lower
     cross. All parameters in the extended hybrid model were frozen except for the blackbody
     temperature ($kT$) and power-law photon index when the XSPEC \emph{steppar} procedure was
     used. The parameter estimation was performed on ungrouped data. }
  \label{fig:GRB911016_GLAST_ufu_chi2}
\end{figure}

XSPEC was utilized on the data produced by {\it gtobssim} and {\it
GBM tools} by imposing the extended hybrid model, as described by
equation (\ref{eqn:exthybrid}), on the time-integrated
spectra from two NaI detectors, one BGO detector and the LAT
instrument. Both $\chi^2$ and C statistic, a modified version of
Cash statistic \cite[see][]{Cash1979, ArnaudDorman2003}, was used
in our analysis to estimate parameters that covered the first five
seconds from trigger, while only $\chi^2$ statistic was used for
the model test. C statistic give better results than $\chi^2$ at
count rates below 10 counts per bin, but can not, as opposed to $\chi^2$ 
tests, be used to get a "goodness-of-fit". It is therefore 
primarily used for estimation of parameter values. C statistic also
assumes that the error on the counts is pure Poissonian, and
should hence be the preferred statistic when estimating parameter
values based on data with low photon counts. At higher count
rates, both $\chi^2$ and C statistic are expected to give the same
parameter estimates, assuming we are dealing with Gaussian
distributions. The $\chi^2$ model analysis was performed on both
ungrouped and grouped data. In the latter case, the data was
grouped into bins with at least 10 photons per bin for all
detectors. Some of the results, from our analysis using $\chi^2$
and C statistic, are presented in table \ref{tab:glastfit}.
\begin{table}
\caption{Parameter values from hybrid model fit on simulated GLAST data}
\label{tab:glastfit}
\begin{tabular}{ccccccccc}
\hline
\tablehead{1}{c}{b}{Burst}
  & \tablehead{1}{c}{b}{Statistic\tablenote{$\chi^2_u$ indicates that $\chi^2$ model
  fit and parameter estimation was performed on ungrouped data, $\chi^2_g$ that the
  same procedure was performed on grouped data with at least 10 photons per bin while
  C indicates that C statistic was used to estimate the parameter values.}}
  & \tablehead{1}{c}{b}{Blackbody}
  & \tablehead{1}{c}{b}{Powerlaw}
  & \tablehead{1}{c}{b}{Powerlaw}
  & \tablehead{3}{c}{b}{Model Statistics}\\
    \tablehead{1}{c}{b}{}
  & \tablehead{1}{c}{b}{}
  & \tablehead{1}{c}{b}{Temp [keV]}
  & \tablehead{1}{c}{b}{Index 1}
  & \tablehead{1}{c}{b}{Index 2}
  & \tablehead{1}{c}{b}{$\nu$}
  & \tablehead{1}{c}{b}{$\chi^2_\nu$}
  & \tablehead{1}{c}{b}{$P$}\\
\hline
 GRB911016   & $\chi^2_u$ & $53.3 \pm 1.1$ & $1.30 \pm 0.04$ & $1.71 \pm 0.03$ & 336 & 0.90 & 0.92 \\
             & $\chi^2_g$ & $53.8 \pm 1.1$ & $1.25 \pm 0.04$ & $1.76 \pm 0.03$ & 206 & 0.98 & 0.55 \\
             & C          & $53.9 \pm 1.1$ & $1.26 \pm 0.04$ & $1.74 \pm 0.03$ &  -  &   -  &  -   \\
         \hline
 GRB941026   & $\chi^2_u$ & $59.2 \pm 1.3$ & 1.6 (frozen)    & 2.1 (frozen)    & 339 & 0.82 & 0.99 \\
             & $\chi^2_g$ & $61.5 \pm 1.3$ & $1.71 \pm 0.03$ & $2.9 \pm 0.8$   & 228 & 0.82 & 0.98 \\
             & C          & $61.0 \pm 1.3$ & $1.69 \pm 0.02$ & $2.4 \pm 0.2$   &  -  &   -  &  -   \\
         \hline
 GRB960530   & $\chi^2_u$ & $41.8 \pm 0.8$ & $1.48 \pm 0.02$ & $2.4 \pm 0.2$   & 336 & 0.91 & 0.87 \\
             & $\chi^2_u$ & $41.8 \pm 0.8$ & $1.42$ (frozen) & 2.1 (frozen)    & 339 & 0.96 & 0.70 \\
             & $\chi^2_g$ & $41.7 \pm 0.9$ & $1.44 \pm 0.02$ & $2.3 \pm 0.1$   & 208 & 1.03 & 0.36 \\
             & C          & $41.5 \pm 0.9$ & $1.42 \pm 0.01$ & $2.2 \pm 0.1$   &  -  &   -  &   -  \\
\hline
\end{tabular}
\end{table}
Figures \ref{fig:GRB911016_GLAST_efspec_s} and
\ref{fig:GRB911016_GLAST_efspec} show a simulated photon spectrum,
based on BATSE data for GRB911016, together with the fit of the
extended hybrid model. The left panel in figure
\ref{fig:GRB911016_GLAST_ufu_chi2} shows the $\nu F _\nu$ spectrum
for the same burst. The resulting fit parameters were, for each
simulated burst, compared with the SBM input parameters. The
simulated GRB911016 output parameters was consistent with the SBM
input parameter values. Table \ref{tab:glastfit} presents some of
the parameter values and also the reduced $\chi^2$- ($\nu$
represents the degrees of freedom) and $P$-values from the hybrid
model fits. All the fits showed reduced $\chi^2$ values around 1.0
and with high $P$-values, close to 1.0, for the ungrouped
data. The high $P$-values indicate that the errors for the ungrouped
data points, used in our XSPEC fits, are over-estimated. Grouping
the data into at least 10 photons per bin, decreased the $P$-values
and also gave $\chi^2$-statistic parameter estimates closer to the
values found when using C-statistic. The right panel in figure
\ref{fig:GRB911016_GLAST_ufu_chi2} shows the $\chi^2$ confidence
regions (with $1\sigma$, $2\sigma$ and $3\sigma$ levels) for the
power-law photon-index versus the blackbody temperature $kT$. It
shows a consistency between the BATSE data and the simulated GLAST
data for the modeled temperature. The temperature was also
consistent with the input data for the other two bursts, but the
mean value for the photon indices became slightly softer in the
GLAST data fits. This is probably due to the evolution of the
blackbody temperature and, in the case of GRB941026, the evolution
of the break-energy of the power-law. Freezing the indices to the
expected values, when analyzing the ungrouped data, we still
received good statistics $\chi^2 \lesssim 1.0$, and consistency
with the input parameters. This can be seen for GRB960530 in table
\ref{tab:glastfit}. It was however not possible to determine the
confidence levels for the high-energy cut-off parameters in the
fit -- neither with binned $\chi^2$ nor C statistic -- due to the
low photon count at the higher energies. \label{sec:results}

\section{Discussion}
\label{sec:discussion} The hybrid model, that may be interpreted
as a combination of photospheric and optically-thin synchrotron
radiation, fits the spectra of single-pulsed GRBs well \citep[see][]{Ryde2005,Battelino2006}. The 
problem with the synchrotron "line of death"
($\alpha\,>\,-$2/3) is avoided with this model, since the Rayleigh-Jeans 
portion of the blackbody component explains the hard spectral slopes seen
in many BATSE bursts.  We presented here simulations of thermal 
bursts with a large super-MeV emission in the GLAST energy range using 
\emph{gtobssim} and \emph{GBM tools}. The simulations were 
followed by XSPEC analysis of the first five seconds of the resulting 
data for each burst, grouped to at least 10 photons per bin. This analysis 
showed that an applied thermal model, consisting of a blackbody function and 
a broken power-law with a high-energy cutoff, gave the desired 
"goodness-of-fit" represented by the $P$-value close to 0.5 for all bursts. 
Parameter estimates of the power-law indices generally gave softer 
values than used for the simulation, but this can be explained by the 
temporal evolution of the blackbody peak and power-law break over the 
five seconds that were integrated into one spectrum.
The simulations of GLAST data for thermal bursts hence show that a 
photospheric component should be clearly detected by the GLAST instruments, 
if it dominates in the energy window of the GBM instrument and is super-positioned 
over an optically-thin synchrotron "background" spectrum extending from the lowest GBM 
energies into the LAT domain. The evolution of the Planck function will be 
possible to determine using events collected by the two detector types, NaI 
(8 keV - 1 MeV) and BGO (150 keV - 30 MeV), that constitute the GBM instrument. The 
non-thermal component will be detectable by both GBM and LAT.

From the simulated GLAST data of the three single-pulse bursts it
was not possible to determine the 1$\sigma$ confidence region for
the high-energy cut-off parameters due to the low photon count,
not even when using C statistic. We therefore expect that the
cut-off and $e$-folding energy for some bursts may be hard to
determine, especially for time-resolved spectra, due to the low
photon count, if the LAT instrument response function is similar
to the one used for Data Challenge 2 (DC2), even if the cut-off
lies within the LAT domain. Possibly better results could be
reached using the raw event data or using a response function with
more lean cuts than the ones used in DC2 for time-integrated
spectra.


\begin{theacknowledgments}
We would like to thank Valerie Connaughton and David Band for
making the GBM Tools software package available and also the GLAST LAT 
team for the ScienceTools package source code. Support for this
work was given by the Swedish National Space Board.
\end{theacknowledgments}

\bibliographystyle{aipproc}   
\bibliography{sim_thermal_grbs.bib}

\begin{thebibliography}{24}
\expandafter\ifx\csname natexlab\endcsname\relax\def\natexlab#1{#1}\fi
\providecommand{\enquote}[1]{``#1''}
\expandafter\ifx\csname url\endcsname\relax
  \def\url#1{\texttt{#1}}\fi
\expandafter\ifx\csname urlprefix\endcsname\relax\def\urlprefix{URL }\fi
\providecommand{\eprint}[2][]{\url{#2}}

\bibitem[Band et~al.(1993)]{Bandetal1993}
D.~Band, et~al., \emph{\apj} \textbf{413}, 281--292 (1993).

\bibitem[{Crider} et~al.(1997)]{Crideretal1997}
A.~{Crider}, E.~P. {Liang}, I.~A. {Smith}, R.~D. {Preece}, M.~S. {Briggs},
  G.~N. {Pendleton}, W.~S. {Paciesas}, D.~L. {Band}, and J.~L. {Matteson},
  \emph{\apjl} \textbf{479}, L39+ (1997).

\bibitem[Preece et~al.(1998)]{Preeceetal1998}
R.~D. Preece, et~al., \emph{\apj} \textbf{506}, L23--L26 (1998).

\bibitem[{Lloyd} and {Petrosian}(2000)]{LloydPetrosian2000}
N.~M. {Lloyd}, and V.~{Petrosian}, \emph{\apj} \textbf{543}, 722--732 (2000),
  \eprint{astro-ph/0007061}.

\bibitem[{Panaitescu} and {M{\'e}sz{\'a}ros}(2000)]{PanaitescuMeszaros2000}
A.~{Panaitescu}, and P.~{M{\'e}sz{\'a}ros}, \emph{\apjl} \textbf{544}, L17--L21
  (2000), \eprint{astro-ph/0009309}.

\bibitem[{Medvedev}(2000)]{Medvedev2000}
M.~V. {Medvedev}, \emph{\apj} \textbf{540}, 704--714 (2000),
  \eprint{astro-ph/0001314}.

\bibitem[{Baring} and {Braby}(2004)]{BaringBraby2004}
M.~G. {Baring}, and M.~L. {Braby}, \emph{\apj} \textbf{613}, 460--476 (2004),
  \eprint{astro-ph/0406025}.

\bibitem[{M{\'e}sz{\'a}ros} and {Rees}(2000)]{MeszarosRees2000}
P.~{M{\'e}sz{\'a}ros}, and M.~J. {Rees}, \emph{\apj} \textbf{530}, 292--298
  (2000), \eprint{astro-ph/9908126}.

\bibitem[{M{\'e}sz{\'a}ros} et~al.(2002)]{Meszarosetal2002}
P.~{M{\'e}sz{\'a}ros}, E.~{Ramirez-Ruiz}, M.~J. {Rees}, and B.~{Zhang},
  \emph{\apj} \textbf{578}, 812--817 (2002), \eprint{astro-ph/0205144}.

\bibitem[{Daigne} and {Mochkovitch}(2002)]{DaigneMochkovitch2002}
F.~{Daigne}, and R.~{Mochkovitch}, \emph{\mnras} \textbf{336}, 1271--1280
  (2002), \eprint{astro-ph/0207456}.

\bibitem[{Drenkhahn} and {Spruit}(2002)]{DrenkhahnSpruit2002}
G.~{Drenkhahn}, and H.~C. {Spruit}, \emph{\aap} \textbf{391}, 1141--1153
  (2002), \eprint{astro-ph/0202387}.

\bibitem[{Ryde}(2004)]{Ryde2004a}
F.~{Ryde}, \emph{\apj} \textbf{614}, 827--846 (2004).

\bibitem[{Ryde}(2005)]{Ryde2005}
F.~{Ryde}, \emph{\apjl} \textbf{625}, L95--L98 (2005),
  \eprint{astro-ph/0504450}.

\bibitem[{Rees} and {M{\'e}sz{\'a}ros}(2005)]{ReesMeszaros2005}
M.~J. {Rees}, and P.~{M{\'e}sz{\'a}ros}, \emph{\apj} \textbf{628}, 847--852
  (2005), \eprint{astro-ph/0412702}.

\bibitem[{Ryde} et~al.(2006)]{Rydeetal2006}
F.~{Ryde}, C.-I. {Bj{\"o}rnsson}, Y.~{Kaneko}, P.~{M{\'e}sz{\'a}ros},
  R.~{Preece}, and M.~{Battelino}, \emph{\apj} \textbf{652}, 1400--1415 (2006),
  \eprint{astro-ph/0608363}.

\bibitem[{Ghirlanda} et~al.(2003)]{Ghirlandaetal2003}
G.~{Ghirlanda}, A.~{Celotti}, and G.~{Ghisellini}, \emph{\aap} \textbf{406},
  879--892 (2003).

\bibitem[Arnaud and Dorman(2003)]{ArnaudDorman2003}
K.~Arnaud, and B.~Dorman, \emph{Xspec, an X-Ray Spectral Fitting Package,
  User's Guide for version 11.3.x}, NASA, 2003.

\bibitem[{Gonz{\'a}lez} et~al.(2003)]{Gonzalezetal2003}
M.~M. {Gonz{\'a}lez}, B.~L. {Dingus}, Y.~{Kaneko}, R.~D. {Preece}, C.~D.
  {Dermer}, and M.~S. {Briggs}, \emph{Nature} \textbf{424}, 749--751 (2003).

\bibitem[Battelino(2006)]{Battelino2006}
M.~Battelino, {Simulation of the keV-GeV emission from gamma-ray bursts using a
  thermal emission model and its detectability by GLAST} (2006), {Master
  Thesis, Department of Astronomy, Stockholm University}.

\bibitem[{de Jager} et~al.(1996)]{deJageretal1996}
O.~C. {de Jager}, A.~K. {Harding}, P.~F. {Michelson}, H.~I. {Nel}, P.~L.
  {Nolan}, P.~{Sreekumar}, and D.~J. {Thompson}, \emph{\apj} \textbf{457},
  253--+ (1996).

\bibitem[{Kocevski} et~al.(2003)]{Kocevskietal2003a}
D.~{Kocevski}, F.~{Ryde}, and E.~{Liang}, \emph{\apj} \textbf{596}, 389--400
  (2003).

\bibitem[{Ryde} and {Svensson}(2000)]{RydeSvensson2000}
F.~{Ryde}, and R.~{Svensson}, \emph{\apjl} \textbf{529}, L13--L16 (2000).

\bibitem[{Ryde} and {Svensson}(2002)]{RydeSvensson2002}
F.~{Ryde}, and R.~{Svensson}, \emph{\apj} \textbf{566}, 210--228 (2002).

\bibitem[{Cash}(1979)]{Cash1979}
W.~{Cash}, \emph{\apj} \textbf{228}, 939--947 (1979).

\end{thebibliography}

\end{document}